\documentclass[aps,prl,twocolumn,showpacs,superscriptaddress,groupedaddress]{revtex4}  
\usepackage{graphicx}  
\usepackage{dcolumn}   
\usepackage{bm}        
\usepackage{amssymb}   

\newcommand{\eq}{\begin{eqnarray}}
\newcommand{\en}{\end{eqnarray}}

\begin{document}

\title{Spectrum of three-body bound states in a finite volume}



\author{Ulf-G. Mei{\ss}ner}

\affiliation{Helmholtz-Institut f\"ur Strahlen- und Kernphysik (Theorie) and
Bethe Center for Theoretical Physics,
Universit\"at Bonn,
D-53115 Bonn, Germany
}

\affiliation{Institute for Advanced Simulation (IAS-4), 
Institut f\"ur Kernphysik 
(IKP-3) and\\ J\"ulich Center for Hadron Physics,
Forschungszentrum J\"ulich, D-52425 J\"ulich, Germany}

\author{Guillermo R\'{\i}os}
\author{Akaki Rusetsky}

\affiliation{Helmholtz-Institut f\"ur Strahlen- und Kernphysik (Theorie) and
Bethe Center for Theoretical Physics,
Universit\"at Bonn,
D-53115 Bonn, Germany
}

\begin{abstract}
The spectrum of a bound state of three identical particles with a mass $m$
in a finite cubic box is studied. 
It is shown that in the unitary limit, the energy shift of a shallow bound 
state is given by $\Delta E=c (\kappa^2/m)\,(\kappa L)^{-3/2}|A|^2
\exp(-2\kappa L/\sqrt{3})$, 
where $\kappa$ is the bound-state momentum,
$L$ is the box size, $|A|^2$ denotes the three-body analog
of the asymptotic normalization
coefficient of the bound state wave function
 and $c$ is a numerical constant. The formula is valid for $\kappa L\gg 1$.

\end{abstract}



\maketitle

\section{Introduction}

Strong interactions between two particles can be studied in
{\em ab initio} lattice simulations, like for hadron-hadron scattering
in Quantum Chromodynamics or di\-mer-di\-mer scattering at ultracold 
temperatures. 
At present, L\"uscher's approach~\cite{luescher-torus}
represents a standard way to study  two-body scattering observables 
on the lattice. In its original form, 
this approach relates the two-particle scattering phase in the
elastic region to the measured energy spectrum of the 
Hamiltonian in a finite volume. In the literature, one finds 
different generalizations of the
L\"uscher approach.  For instance, the approach has been formulated
in case of moving frames~\cite{moving}, (partially) twisted boundary 
conditions~\cite{PT} and 
for coupled-channel scattering~\cite{coupled-channel} (for a recent application
of this approach to the analysis of the two-channel case on the lattice,
see Ref.~\cite{jlab}). A closely related 
framework based on the use of the unitarized ChPT in a finite volume has been 
also proposed~\cite{unitary-ChPT}. Further, a method for the measurement
of  resonance matrix elements and form factors in the time-like region
has been worked out~\cite{matrixelements}. Note, however that all these 
generalizations explicitly deal with  two-body channels. Studying a 
genuinely three-body problem in a finite volume has proven to be a far more
complicated enterprise and, 
albeit there have been several attempts to solve this problem
in the last few years~\cite{Polejaeva,Sharpe,BD,Peng,DeanLee,kreuzer}, 
the method is  still in its infancy. On the other hand, recent 
progress on the lattice, related to the study of the inelastic 
resonances such as the Roper resonance~\cite{Roper}, and of 
the properties of light nuclei~\cite{NPLQCD,Yamazaki,nuclear-EFT},
indicates that the generalization of the L\"uscher method to the multi-particle
(three and more) systems is urgently needed.

The main obstacle that one encounters in generalizing L\"uscher's approach from
two to three particles has a transparent physical interpretation. In the 
center-of-mass (CM) frame, the two-body scattering can be considered as a 
scattering of one particle in a given potential. If this potential has a 
short range (much smaller than the box size $L$), then the scattering wave 
function at the boundaries will depend only on the scattering phase shift
in the infinite volume and, therefore, the discrete spectrum in a finite box
will be determined by this phase shift only. In other words, the spectrum in
a large but finite box does not depend on the details of the interaction at 
short distances. This is not so obvious in case of three particles. In this 
case, each pair of particles can come close to each other and be still
separated from the third one by a large distance of order $L$. It took 
a certain effort to prove that, despite the fact that such configurations 
are allowed,
the finite-volume spectrum is still determined solely by the 
infinite-volume $S$-matrix elements and does not depend on the short-range
details of the interaction~\cite{Polejaeva}, see also Refs.~\cite{BD,Sharpe}. 
For instance, in a recent paper~\cite{Sharpe} the authors succeeded in deriving
a quantization condition for the three-particle spectrum in a finite volume.
It has a quite complicated structure, in particular, due to the fact that the
infinite-volume amplitudes that enter this condition are defined in a 
unconventional manner (the necessity of such a definition has been pointed
out already in Ref.~\cite{Polejaeva}). For this reason, it is not an easy task
to use this quantization condition for the analysis of lattice data
-- in fact, we are not aware of a single explicit prediction for the 
volume dependence of physical observables except for the 
ground-state shift of identical particles~\cite{groundstate}, 
which were done in this 
formalism so far~\footnote{In this respect we would like to note that 
the results of  Refs.~\cite{BD,DeanLee} are obtained in the context of a 
particle-dimer bound-state problem.}. Note also that in Ref.~\cite{kreuzer},
in the framework of the non-relativistic EFT, 
it has been explicitly demonstrated that carrying out the renormalization
in the infinite volume leads to the cutoff-independent three-particle 
bound-state spectrum in a finite volume that is equivalent to the statement
that this spectrum is determined by the $S$-matrix elements in the infinite
volume.

The aim of the present paper is to obtain such an explicit volume dependence
for the physical quantity which, in our opinion, is the easiest to handle.
In particular, we consider  shallow bound states of three identical particles
in the unitary limit. This means that the two-body scattering length $a$ tends
to infinity and the corresponding effective range is zero. The three-body 
bound-state momentum $\kappa$, which is related
to the binding energy $E_T$ through $E_T=\kappa^2/m$, is much smaller than
the particle mass $m$ or the inverse of the interaction range. Still, we 
consider large boxes where $\kappa L\gg 1$. Our treatment of the three-body
bound state is not based on the quantization condition derived in 
Ref.~\cite{Sharpe}, but closely follows the two-body pattern of 
Ref.~\cite{luescher-stable} (see also~\cite{DeanLee,koenig}, where, 
in particular, the result of Ref.~\cite{luescher-stable} is generalized 
to the case of an arbitrary angular momentum). For this reason,
our explicit result provides a beautiful testing ground for the general 
approach formulated in Ref.~\cite{Sharpe} and helps to better understand 
its structure. On the other hand, our result can be immediately verified 
through numerical calculations in a finite volume similarly 
to those carried out, e.g., in Ref.~\cite{kreuzer} that provides an additional
check on the theoretical framework.

\section{Derivation of the formula for energy shift}

We start from the Schr\"odinger equation for three identical particles
in the infinite volume
\eq
\biggl\{\sum_{i=1}^3\biggl(-\frac{1}{2m}\,\nabla_i^2+V({\bf x}_i)\biggr)
+E_T\biggr\}
\psi({\bf r}_1,{\bf r}_2,{\bf r}_3)=0\, ,
\en
where $\nabla_i=\partial/\partial {\bf r}_i$ and 
the Jacobi coordinates are defined as
\eq
{\bf x}_i={\bf r}_j-{\bf r}_k\, ,\quad\quad
{\bf y}_i=\frac{1}{\sqrt{3}}\,({\bf r}_j+{\bf r}_k-2{\bf r}_i)\, ,
\en
with $(ijk)=(123),(312),(231)$. Here, for simplicity, we assume that no
three-body force is present. The inclusion of the latter can be done 
in analogy with Ref.~\cite{Polejaeva}.

In a finite volume, the potential $V$ is replaced by a sum over all mirror
images
\eq
V_L({\bf x}_i)=\sum_{{\bf n}\in\mathbb{Z}^3}V({\bf x}_i+{\bf n}L)\, ,
\en
and the Schr\"odinger equation 
takes the form
\eq
\biggl\{\sum_{i=1}^3\biggl(-\frac{1}{2m}\,\nabla_i^2+V_L({\bf x}_i)\biggr)
+E_L\biggr\}
\psi_L({\bf r}_1,{\bf r}_2,{\bf r}_3)=0\, .
\en
In the CM frame, the bound-state wave functions $\psi,\psi_L$ depend on two
Jacobi coordinates ${\bf x}_i,{\bf y}_i$. For three identical particles,
$\psi({\bf x}_i,{\bf y}_i)=\psi({\bf x}_k,{\bf y}_k)\, ,~i,k=1,2,3,$ 
and similarly
to the finite-volume wave function $\psi_L$.

In order to evaluate the finite-volume shift $\Delta E=E_T-E_L$,
in analogy to Ref.~\cite{luescher-stable}, we define in the CM frame the 
trial wave function (we choose $i=1$ from now on)
\eq
\psi_0
=\sum_{{\bf n},{\bf m}}
\psi\biggl({\bf x}_1-({\bf n}+{\bf m})L,{\bf y}_1+\frac{1}{\sqrt{3}}\,
({\bf n}-{\bf m})L\biggr)\, .
\en
Denoting $H_L=\sum_{i=1}^3\biggl(-\frac{1}{2m}\,\nabla_i^2+V_L({\bf x}_i)\biggr)$, it can be straightforwardly checked that $\psi_0$ obeys the equation
$(H_L+E_T)\psi_0=\eta$, where
\eq\label{eq:eta}
\eta=\sum_{{\bf n},{\bf m}}\hat V_{\bf nm}
\psi\biggl({\bf x}_1-({\bf n}+{\bf m})L,{\bf y}_1+\frac{1}{\sqrt{3}}\,
({\bf n}-{\bf m})L\biggr)
\en
and
\eq
\hat V_{\bf nm}&=&
\sum_{{\bf k}\neq-{\bf n}-{\bf m}}V({\bf x}_1+{\bf k}L)
+\sum_{{\bf k}\neq {\bf n}}V({\bf x}_2+{\bf k}L)
\nonumber\\
&+&\sum_{{\bf k}\neq {\bf m}}V({\bf x}_3+{\bf k}L)\, .
\en
Since the potential $V({\bf x})$ has a short range,
the quantity $\eta$ exponentially vanishes at a large $L$,
$\eta\propto \exp(-{\rm const}\cdot \kappa L)$.
Further, applying perturbation theory, it can be
 verified that, to all orders, the energy shift is given by 
\eq
\Delta E&\!\!=\!\!&\frac{\langle\psi_0|T|\psi_0\rangle}{\langle\psi_0|\psi_0\rangle}\, ,
\nonumber\\
T&\!\!=\!\!&(H_L+E_T)-(H_L+E_T)QGQ(H_L+E_T)\, ,
\en
where
\eq
G=\frac{1}{H_L+E_L}\, ,\quad\quad
Q=\frac{|\psi_0\rangle\langle\psi_0|}{\langle\psi_0|\psi_0\rangle}\, .
\en
Since the quantity $\eta$ is exponentially suppressed at a large $L$, the 
leading exponential correction to the energy shift is given by (cf. with 
Ref.~\cite{luescher-stable})
\eq
\Delta E=\frac{\langle\eta|\psi_0\rangle}{\langle\psi_0|\psi_0\rangle}
+\cdots\, .
\en
Note that in Ref.~\cite{DeanLee} this formula in case of more than two particles was given without derivation. 
A detailed derivation in case of two particles is given in~\cite{koenig}.

In the next step, one substitutes the expression for $\eta$ from 
Eq.~(\ref{eq:eta}) into the above expression for the energy shift and 
picks those terms that give a leading exponential contribution at large $L$. 
Taking into account the fact that the argument of the exponent in the 
infinite-volume  wave function $\psi({\bf x}_1,{\bf y}_1)$ is proportional 
to the hyperradius $R=\frac{1}{\sqrt{2}}\,({\bf x}_1^2+{\bf y}_1^2)^{1/2}$, 
one has to minimize the sum of two hyperradii, coming from two wave functions 
in the overlap integral. Finally, the expression of the energy shift at 
leading order takes the form
\eq\label{eq:master}
\Delta E&\!\!=\!\!&6\cdot 2\cdot 3\int d^3{\bf x}_1d^3{\bf y}_1 
\psi({\bf x}_1,{\bf y}_1)V({\bf x}_1)
\nonumber\\
&\!\!\times\!\!&\psi\biggl({\bf x}_1-{\bf e}L,{\bf y}_1+\frac{1}{\sqrt{3}}\,{\bf e}L\biggr)+\cdots\, ,
\en
where ${\bf e}=(0,0,1)$ denotes a unit vector and the ellipses stand for the
exponentially suppressed terms. In this formula, the infinite-volume wave 
function $\psi$ is normalized to unity. The factor in front of the integral 
reflects the symmetries: 6 for different orientations of the unit vector 
${\bf e}$, 2 for different signs in the second argument of the wave function
${\bf y}_1\pm \frac{1}{\sqrt{3}}\,{\bf e}L$, and 3 for three different 
pair potentials.

\section{Evaluation of the energy shift}

In order to evaluate the overlap integral that defines the leading order energy
shift, an explicit expression for the bound-state wave function should be 
supplied. Only the asymptotic tail of the wave function matters, since the 
finite-volume spectrum is uniquely determined by the $S$ matrix elements 
in the infinite volume~\cite{Polejaeva}. Here, we shall be working in the 
unitary limit. In the context of the lattice this means that the two-body
scattering length $a\geq L$, i.e., even at the box boundaries, the hyperradius
$R\leq a$. On the other hand, we assume that the interaction range is much 
smaller than $L$. Under these assumptions, almost everywhere in the 
configuration space, the wave function can be approximated by the
well-known universal expression (see, e.g.~\cite{BraatenHammer})
\eq\label{eq:psi}
\psi({\bf x}_1,{\bf y}_1)&\!\!=\!\!&A{\cal N}R^{-5/2}f_0(R)\sum_{i=1}^3
\frac{\sinh(s_0(\pi/2-\alpha_i))}{\sin(2\alpha_i)}
\nonumber\\
&\!\!\doteq\!\!&\sum_{i=1}^3\phi(R,\alpha_i)\, ,
\en
where
\eq
f_0(R)=R^{1/2}K_{is_0}(\sqrt{2}\kappa R)
\en
and $K_\nu(z)$ denotes the Bessel function. Here, 
$\alpha_i=\arctan(|{\bf x}_i|/|{\bf y}_i|)$ are Delves hyperangles 
and the numerical constant $s_0\simeq 1.00624$ is the solution to the 
transcendental equation
\eq
s_0\cosh\frac{\pi s_0}{2}=\frac{8}{\sqrt{3}}\,\sinh\frac{\pi s_0}{6}\, .
\en
Further, ${\cal N}$ is the normalization coefficient of the exact asymptotic 
wave function in Eq.~(\ref{eq:psi}), so that
\eq
\int d^3{\bf x}_1 d^3{\bf y}_1 |\psi({\bf x}_1,{\bf y}_1)|^2=|A|^2\, .
\en
Evaluating the integral explicitly, one gets
\eq
{\cal N}^2&\!\!=\!\!&\kappa^2C_0\, ,
\nonumber\\
C_0^{-1}&\!\!=\!\!&\frac{8\pi^3}{\sinh(\pi s_0)}\,
\biggl(\frac{3}{4}\,\sinh(\pi s_0)
-\frac{3\pi s_0}{4}
\nonumber\\
&\!\!-\!\!&\frac{4\pi}{\sqrt{3}}\,\sinh\frac{\pi s_0}{3}
+\frac{2\pi}{\sqrt{3}}\,\sinh\frac{2\pi s_0}{3}\biggr)\, .
\en
Finally, the quantity $|A|^2$ denotes a three-body analog of
the asymptotic normalization coefficient for the wave function.
It encodes the information about
the short-range dynamics in the system. Namely, if in the creation of the bound
state the long-range effects dominate, it is expected that
the quantity $|A|^2$  is close to
one~\footnote{This is an analog of Weinberg's compositeness 
condition~\cite{Weinberg} in case of three particles. The 
compositeness can be studied on the lattice, see, 
e.g.,~Ref.~\cite{GuoRios}.}.

Next, we evaluate the overlap integral in Eq.~(\ref{eq:master}) by using
the explicit wave function from Eq.~(\ref{eq:psi}). In analogy to 
Eq.~(\ref{eq:psi}), the second wave function in the integral can be written as
\eq
\psi\biggl({\bf x}_1-{\bf e}L,{\bf y}_1+\frac{1}{\sqrt{3}}\,{\bf e}L\biggr)
=\sum_{i=1}^3\phi(R',\alpha'_i)\, .
\en
As $L\to\infty$,
\eq
R'&=&\frac{(({\bf x}_1-{\bf e}L)^2+
({\bf y}_1+{\bf e}L/\sqrt{3})^2)^{1/2}}{\sqrt{2}}\nonumber\\
&\to& \sqrt{\frac{2}{3}}\,L
-\frac{\sqrt{3}}{2\sqrt{2}}\,{\bf e}\cdot{\bf x}_1+\frac{1}{2\sqrt{2}}{\bf e}\cdot{\bf y}_1+\cdots\, ,
\en
whereas the angular variables tend to the following limiting values:
\eq
\tan\alpha_1'&\!\!=\!\!&\frac{|{\bf x}_1-{\bf e}L|}{|{\bf y}_1+{\bf e}L/\sqrt{3}|}\to\sqrt{3}+\cdots\, ,
\nonumber\\
\tan\alpha_2'&\!\!=\!\!&\frac{|{\bf x}_2+{\bf e}L|}{|{\bf y}_2+{\bf e}L/\sqrt{3}|}\to\sqrt{3}+\cdots\, ,
\nonumber\\
\tan\alpha_3'&\!\!=\!\!&\frac{|{\bf x}_3|}{|{\bf y}_3-{2\bf e}L/\sqrt{3}|}\to
\frac{\sqrt{3}}{2}\,\frac{|{\bf x}_3|}{L}+\cdots
\nonumber\\
&\!\!=\!\!&\frac{\sqrt{6}R\sin\alpha_3}{2L}+\cdots\, .
\en
The expansion of the angular part of the second wave function yields
\eq
\sum_{i=1}^3
\frac{\sinh(s_0(\pi/2-\alpha'_i))}{\sin(2\alpha'_i)}\to
\frac{L}{\sqrt{6}R}\,\frac{\sinh(\pi s_0/2)}{\sin(\alpha_3)}+\cdots\, .
\en
Using the Faddeev equation
\eq
&&\hspace*{-.3cm}
\psi({\bf x}_1,{\bf y}_1)V({\bf x}_1)=\biggl(\frac{1}{m}\,\biggl(
\frac{\partial^2}{\partial {\bf x}_i^2}+\frac{\partial^2}{\partial {\bf y}_i^2}
\biggr)-E_T\biggr)\phi(R,\alpha_1)\, ,
\nonumber\\
&&\hspace*{-.3cm}
\en
and taking into account the fact that for the wave function defined in Eq.~(\ref{eq:psi}),
the following relation is valid:
\eq
\biggl(\frac{1}{m}\,\biggl(\frac{\partial^2}{\partial {\bf x}_i^2}+ 
\frac{\partial^2}{\partial {\bf y}_i^2}\biggr)-E_T\biggr)\phi(R,\alpha_1)
\nonumber\\
=-\delta^3({\bf x}_1)f(|{\bf y}_1|)\, ,
\en
where
\eq
f(|{\bf y}_1|)=\frac{4\pi}{m}\,A{\cal N}\,\frac{K_{is_0}(\kappa|{\bf y}_1|)
\sinh\bigl(\frac{\pi s_0}{2}\bigr)}{|{\bf y}_1|}\, ,
\en
the expression for the energy shift simplifies to
\eq
\Delta E=-36\int d^3{\bf y}_1\,f(|{\bf y}_1|)\psi\biggl(-{\bf e}L,{\bf y}_1
+\frac{1}{\sqrt{3}}\,{\bf e}L\biggr)\, .
\en
Next, using the asymptotic expression for the hyperradial wave function
\eq
f_0(R')&\to&\sqrt{\frac{\pi}{2}}\exp\biggl(-\frac{2\kappa L}{\sqrt{3}}\biggr)
\frac{1}{(\sqrt{2}\kappa)^{1/2}}
\nonumber\\[2mm]
&\times&
\exp\biggl(\frac{\sqrt{3}\kappa}{2}{\bf e}\cdot{\bf x}_1-\frac{\kappa}{2}\,{\bf e}\cdot{\bf y}_1\biggr)+\cdots\, ,
\en
we arrive at the final result for the energy shift
\eq\label{eq:final}
\Delta E=c (\kappa^2/m)\,(\kappa L)^{-3/2}|A|^2
\exp(-2\kappa L/\sqrt{3})+\cdots\, ,
\en
where
\eq
c=-36\cdot 3^{3/4}\pi^{7/2}C_0 \sinh^2(\pi s_0/2)\simeq - 96.351\, ,
\en
and the ellipses stand for the sub-leading terms in $L$, both exponentially 
and power-suppressed ones. Note that this behavior qualitatively agrees with 
the result given in Refs.~\cite{kreuzer},
albeit a more detailed numerical study of the problem is needed.

The Eq.~(\ref{eq:final}) is the main result of this paper. 
Measuring the binding energy at different volumes, one may determine 
the infinite-volume quantities $E_T$  and $|A|^2$ 
through the extrapolation procedure.

\section{Conclusions}

The equation~(\ref{eq:final}) is an explicit
prediction of the volume dependence for a genuine three-body observable.
This dependence can be readily verified by using numerical methods that
represents a highly non-trivial check of the whole approach. 
Moreover, 
understanding the present result in the more general context of the three-body
quantization condition, one may gain insight into the complicated three-body
formalism. In view of the recent progress in the study of  inelastic resonances
and nuclei in lattice QCD, this kind of information will be very important.

As mentioned above, the present result is valid within certain approximations.
In the future, we plan to go beyond these approximations.
For example, the next step could be to study the effects of the 
partial-wave mixing in the three-particle
systems as well as the effects of a finite scattering length and
interaction range. Further, it would be extremely interesting (and much 
more challenging) to address the observables from the scattering sector as well.

\begin{acknowledgments}
The authors would like to thank S.~Bour, M.~D\"oring, E.~Epelbaum, H.-W.~Hammer, M.~Hansen, 
M.~Jansen, D.~Lee and S.~Sharpe for useful discussions.
This work is partly supported by the EU
Integrated Infrastructure Initiative HadronPhysics3 Project  under Grant
Agreement no. 283286. We also acknowledge the support by the DFG (CRC 16,
``Subnuclear Structure of Matter'' and CRC 110, ``Symmetries and the Emergence of Structure in QCD'') and by
the Shota Rustaveli National Science Foundation
(Project DI/13/02).
This research is supported in part by Volkswagenstiftung
under contract no. 86260.
\end{acknowledgments}

\end{document}